# The E.U.'s Artificial Intelligence Act: An Ordoliberal Assessment

Manuel Wörsdörfer[1]

**Abstract:** In light of the rise of generative AI and recent debates about the socio-political implications of large-language models and chatbots, this article investigates the E.U.'s Artificial Intelligence Act (AIA), the world's first major attempt by a government body to address and mitigate the potentially negative impacts of AI technologies. The article critically analyzes the AIA from a distinct economic ethics perspective, i.e., 'ordoliberalism 2.0' – a perspective currently lacking in the academic literature. It evaluates, in particular, the AIA's ordoliberal strengths and weaknesses and proposes reform measures that could be taken to strengthen the AIA.

## 1. Introductory Remarks

The rise of generative AI – including chatbots such as ChatGPT and Bard and the underlying large-language models (EPRS, 2023; Floridi, 2023) – has sparked debates about the socio-economic and political implications of AI. Critics point out that these technologies could exacerbate the spread of disinformation, deception, fraud, and manipulation, amplify discrimination risks and biases (thereby negatively affecting the human rights of vulnerable and marginalized populations), trigger hate crimes and oppression, lead to mass surveillance, and further undermine trust in democracy and the rule of law (Ajayi et al., 2023; Bender et al., 2021; Center for AI Safety, 2023b; Gebru et al., 2023a, 2023b; The Guardian, 2023a).[2]

As these discussions show, there appears to be a growing need for (some form of) AI regulation, and several governments have already begun taking initial measures. The Competition and

---

[1] *Address for correspondence:* Dr. Manuel Wörsdörfer, Assistant Professor of Management and Computing Ethics, Maine Business School & School of Computing and Information Science; University of Maine; E-Mail: manuel.woersdoerfer@maine.edu.
*Acknowledgment*: The author would like to thank two anonymous reviewers for their invaluable constructive feedback and criticism. They helped to improve the article significantly. The usual caveats apply.
[2] Some critics have backed their criticism with (a call for) action: Hinton, for example, has resigned from Google and warns of the dangers of AI (New York Times, 2023); Musk et al. have signed an open letter demanding the pause of AI development (Future of Life Institute, 2023; see Gebru et al., 2023); Altman calls for AI (self-)regulation to mitigate the 'risks of increasingly powerful AI' (The Guardian, 2023b); the Center for AI Safety released a statement which warns that AI poses a severe 'risk of extinction' and could be as deadly as pandemics and nuclear weapons (the statement received widespread support from leading AI companies and scientists) (Center for AI Safety, 2023a, 2023b); and other researchers and politicians argue that global governance and an international AI agency are needed (i.e., 'IAEA for AI') (Chowdhury, 2023; Marcus & Reuel, 2023; United Nations, 2023).

Markets Authority (2023) in the U.K., for instance, has launched an investigation examining 'AI foundation models'; the White House (2022) has released an AI Bill of Rights draft, and VP Harris has met with several leaders of big tech (White House 2023a, 2023b); and OpenAI's CEO has testified before Congress. The boldest move so far has been made by the European Commission, which released its Artificial Intelligence Act (AIA) draft proposal in 2021 (European Commission, 2021a, 2021b; see also European Commission, 2020, 2021d, 2022a, 2022b), followed by extensive deliberations and negotiations in the Council of the European Union and European Parliament in 2022, and the approval of the revised draft by the Parliament in 2023.

This article uses a qualitative research method (i.e., a literature review) to analyze the AIA from a distinct economic ethics perspective, i.e., 'ordoliberalism 2.0' (Wörsdörfer, 2020, 2022b). Such a perspective is currently lacking in the academic literature. The paper thus closes a significant research gap and makes important contributions to the predominantly legal-political discourse on the AIA by adding an entirely new (i.e., economic ethics) perspective and a new way of arguing – one that could (ideally) help to make AI systems (more) secure, (more) ethical, and (more) trustworthy (i.e., safety and security are crucial preconditions for gaining the trust of the public; ordoliberalism 2.0. could help companies and government agencies address some of the above ethical concerns while harnessing the advantages of AI technologies, thereby gaining societal trust, disseminating technological advancements, and boosting the economy [for more information, see Wörsdörfer, 2023c]).

The following sections explore, in particular, the Act's institutional strengths and weaknesses and propose ordoliberal-inspired reform measures that could help to 'harden' the AIA. They build on our previous work on ordoliberalism (Wörsdörfer 2013, 2020, 2022c) and the AIA (Wörsdörfer, 2023b). The paper argues, specifically, that the AIA has the potential to realize central ordoliberal principles for the digital economy (Wörsdörfer, 2022b), but for this to happen, key reform steps need to be taken, such as the ones mentioned in Section 5.

The article is structured as follows: Section 2 briefly summarizes crucial ordoliberal principles. Section 3 provides an overview of the critical elements of the AIA by taking a closer look at both the Commission's proposal and the amendments requested by the Council and the Parliament. Section 4 analyzes the AIA's strengths and weaknesses – from a (revised) ordoliberal point of view. Section 5 offers ordoliberal-inspired reform proposals, which could help to further strengthen the AIA. The article ends with a summary of its main findings (Section 6).

## 2. Ordoliberalism

Ordoliberalism is a business-ethical concept that paved the way for Germany's – and the E.U.'s – social market economy implemented after World War II. It attempts to bridge the gap between moral (i.e., social justice, human rights) and economic (i.e., competition, market freedom) imperatives. Its primary goal is to establish an economically efficient and, at the same time, humane socio-economic order – one that can protect the Kantian values of freedom, autonomy, and dignity (Oppenheimer, 1933; Röpke, 1944/1949; Rüstow, 2001; Wörsdörfer, 2013). Two schools of economic thought must be distinguished within classical ordoliberalism – the Freiburg School of Law and Economics and 'sociological neoliberalism.' The founders of the *Freiburg School*, an interdisciplinary research group at the University of Freiburg, were the economist Eucken and the two legal scholars Böhm and Großmann-Doerth; the leading representatives of *sociological neoliberalism* were Rüstow and Röpke (Wörsdörfer, 2022c). Besides classical ordoliberalism, there is also *contemporary* ordoliberalism (Feld & Köhler, 2011; Goldschmidt, 2002, 2007; Goldschmidt & Wohlgemuth, 2008; Häußermann & Lütge, 2022; Vanberg, 2004, 2005, 2008b, 2013; Wörsdörfer, 2020), which bears considerable resemblances to constitutional economics à la Buchanan and evolutionary liberalism à la Hayek.

Amongst the most famous ordoliberal catchwords is the differentiation between *Ordnungspolitik* and *Prozesspolitik* (Eucken, 1950/1965, 1952/2004, 1999, 2001): According to the 'Freiburg imperative,' regulatory or ordering policy is favored, which means that the government as a legislator and rule-maker – and not as a significant economic player – is responsible for setting, preserving, and enforcing the regulatory framework. The government should restrict itself to economic policies that frame or define the general terms and conditions under which market transactions occur. In other words, the government must focus solely on establishing, monitoring, and enforcing the 'rules of the game' instead of steering, influencing, or intervening in market processes and the play itself. The overall goal of regulatory policy is to implement a competitive socio-economic order capable of safeguarding freedom, autonomy and citizen sovereignty, and dignity (this presupposes a constitutional state based on the rule of law).

Eucken's Principles of Economic Policy – and his Constituent and Regulating Principles[3] (Eucken, 1952/2004) – demand not only the disempowerment of socio-economic lobbying and pressure groups; they also require a 'market-conforming' regulatory policy (Röpke, 1942, 1944/1949) – one that does not interfere in the market and price mechanism – and the dismissal of a market-non-conforming process policy (Eucken, 1952/2004).

According to (classical) ordoliberalism, it is also crucial to discuss the relationship between markets and states and clearly define the government's tasks and the limits of its responsibilities. The ideal ordoliberal state is a strong and independent constitutional state (Röpke, 1942, 1944/1949, 1950; Rüstow, 1955), a state that stands above special interest groups and serves as a 'market police' (Röpke, 1942; Rüstow, 1957, 2001), as an 'ordering power,' and as a 'guardian of the competitive order' (Eucken, 1952/2004; Röpke, 1944/1949). The state should ideally be able to fend off special interest groups, keep to the principles of neutrality and impartiality, and confine itself to regulatory policy. The underlying liberal ideals are equality before the law (i.e., the rule of law), freedom of privileges, and the principle of non-discrimination (Böhm, 1966/1980; Vanberg, 2008a) – similar to that of Buchanan's and Hayek's constitutional economics (Brennan & Buchanan, 1985/2000; Buchanan, 1975/2000; Buchanan & Congleton, 1998; Buchanan & Tullock, 1962/1999; Congleton, 2013; Vanberg, 2008b; Wörsdörfer, 2020[4]).

Overall, the ordoliberals are searching for an integrative third way (Oppenheimer, 1933; Rüstow, 2001) between laissez-faire capitalism (i.e., Scylla of paleo-liberalism) and totalitarian collectivism (i.e., Charybdis of socialism). They refer to this type of socio-economic model as ordoliberalism (with 'ordo' simply meaning socio-economic and political order [Eucken, 1950/1965]), social liberalism, or economic humanism (Röpke, 1944/1949; Rüstow, 2001).

Figure 1 summarizes the principles of revised ordoliberalism (i.e., 'ordoliberalism 2.0').

---

[3] Eucken's *Constituent Principles* include 1. competitive market order; 2. primacy of monetary policy/price stability; 3. open markets; 4. private property rights; 5. freedom of contracts; 6. principle of liability; 7. long-term orientation of economic policy; and 8. interdependency of all Constituent Principles (Eucken, 1952/2004). Eucken's *Regulating Principles* include 1. correction of market powers; 2. income redistribution; 3. correction of negative external effects; and 4. correction of 'abnormal supply reactions' (Eucken, 1952/2004; Wörsdörfer, 2022b).

[4] According to Wörsdörfer (2022b), 'ordoliberalism 2.0' rests on the following principles: competitive economy, open markets, freedom of contract/liability, correction of market power, limiting rent-seeking, regulatory and competition policy, rule of law, freedom of privileges/non-discrimination, subsidiarity, and correction of negative external effects.

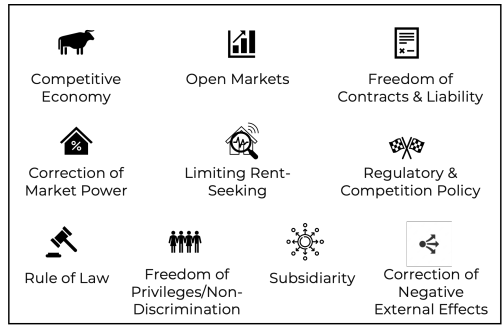

*Figure 1: Principles of 'Ordoliberalism 2.0'*

### 3. Artificial Intelligence Act

The primary goal of the AIA is to create a legal framework for secure, trustworthy, and ethical AI (European Commission, 2021a). That is, it aims to ensure that AI technologies are human-centric, safe to use, compliant with the law, and respectful of fundamental rights, especially those enshrined in the Charter of Fundamental Rights (European Commission, 2021a).

The AIA is the result of extensive stakeholder consultation, incorporates input from the High-Level Expert Group on AI (see its 'Ethics Guidelines for Trustworthy AI' [2019]), and builds on the E.U. White Paper (European Commission, 2020). It is also part of the E.U.'s digital single market strategy and New Legislative Framework (European Commission, n.d.). As such, it complements the General Data Protection Regulation (GDPR) (Greenleaf, 2021) and is consistent with the Digital Services Act (DSA), Digital Markets Act (DMA) (Hacker et al., 2023; Wörsdörfer, 2023a; see Wörsdörfer, 2020, 2021, 2022a), and other regulatory initiatives such as the Machinery Products Regulation, the Data Act and Data Governance Act, the AI Liability Directive, the 'revised Product Liability Directive,' and sectoral product safety frameworks (Dheu et al., 2022; Hacker et al., 2023; Mazzini & Scalzo, 2022).

According to Art. 1, the AIA lays down "harmonized rules for the placing on the market, the putting into service, and the use of [AI] systems […] in the Union; (a) prohibitions of certain [AI] practices; (b) specific requirements for high-risk AI systems and obligations for operators of such systems; (c) harmonized transparency rules for AI systems intended to interact with natural persons, emotion recognition systems and biometric categorization systems, and AI systems used to generate or manipulate image, audio, or video content; (d) rules on market monitoring and surveillance." The regulation applies to "providers of placing on the market or putting into service AI systems in the Union, irrespective of whether those providers are established within the Union or in a third country" (Art. 2). That is, the AIA applies to both private and public actors inside and outside of the E.U. as long as the AI system impacts E.U. citizens. According to the AIA and the Annexes document (European Commission, 2021b), *AI systems* include machine/deep learning, logic- and knowledge-based (expert systems), and statistical (Bayesian) approaches. To ensure technology neutrality and make the definition future-proof, the list can be amended.

The AIA distinguishes between various risk categories – unacceptable, high, and low or minimal risk: That is, it denies market access whenever the risks are deemed too high for risk-mitigating measures (i.e., prohibited AI systems). For high-risk AI systems, market access is granted if they comply with the AIA's ex-ante technical requirements and ex-post market monitoring procedures; minimal-risk systems need to fulfill general safety requirements, i.e., they need to comply with the General Product Safety Directive and sector-specific laws (Almada & Petit, 2023).

*Prohibited AI systems* pose an unacceptable risk or contravene E.U. values, e.g., by violating fundamental rights. Examples include systems that involve the manipulation of people, the exploitation of vulnerabilities, or the causing of physical or psychological harm. All of these 'manipulative, exploitative, and social control practices' are prohibited by the AIA, given that they can potentially be harmful and violate human rights (European Commission, 2021a). The AIA also rules out 'AI-based social scoring for general purposes done by public authorities' and the use of real-time and remote biometric identification systems in publicly accessible spaces for law enforcement purposes (Almeida et al., 2021; European Commission, 2021a). Note that the European Parliament (2023a, 2023b, 2023c) wants to add predictive policing to the list of prohibited technologies and include a complete ban on all AI systems used for emotion recognition and biometric surveillance).

*High-risk AI systems* are technologies that might pose a significant risk to the health and safety of individuals or negatively affect the fundamental rights of persons (European Commission, 2021a). The AIA and Annex III classify the following AI systems as high-risk: Real-time and post remote biometric identification systems, as well as AI systems used for the management and operation of critical infrastructure, education and vocational training, employment and workers management, access to and enjoyment of essential public services and benefits, law enforcement, migration, asylum, and border control management, and the administration of justice and democratic processes (note that the Council's [2022a, 2022b] and Parliament's [EPRS, 2023; European Parliament, 2023a, 2023b, 2023c] proposals want to classify 'general-purpose AI systems,' 'generative foundation models,' and political micro-targeting as high-risk).

Once classified as a high-risk system, system providers need to fulfill several mandatory *requirements*, including risk management system and impact assessment, data governance and management practices (Hacker, 2021), technical documentation, record-keeping and traceability, transparency and information provision, human oversight (Laux, 2023), and accuracy, robustness, and cybersecurity. Providers of high-risk AI systems are also obliged to affix the CE marking their systems (European Commission, 2021a), set up a quality management system, draw up technical documentation, ensure that AI systems undergo conformity assessment procedures, generate AI system logs, take necessary corrective actions if the system is not in conformity, immediately inform national competent authorities, e.g., about changing risk assessment, issues of non-compliance, and corrective actions taken, and cooperate with competent authorities (i.e., provide authorities with necessary information and documentation and grant them access to their logs). The main objective of the conformity assessment procedure is to demonstrate the compliance of high-risk AI systems with the respective AIA requirements. Noteworthy is that high-risk AI systems must undergo a new conformity assessment procedure whenever they are substantially modified. AI systems that successfully pass the procedure will be certified. The certificates will be valid for up to five years and can be renewed – following a re-assessment process – but they can also be suspended or withdrawn. Before marketing them, providers must also register their AI systems in a newly created *E.U.-wide database* for stand-alone high-risk AI systems. The database is operated by the Commission, and the information contained in the database must be accessible to the public.

AI systems considered *low or minimal risk* need to fulfill certain transparency obligations. Such systems include all those interacting with humans, detecting emotions or determining associations with social categories based on biometric data, or generating and manipulating image, audio, or video content. Natural persons need to be informed that they are interacting with the previously mentioned systems (note that based on the Parliament's recommendations, bots and deepfakes would be considered high-risk [European Parliament, 2023a, 2023b, 2023c]).

The AIA also demands the creation of a *European Artificial Intelligence Board* (EAIB). The board will consist of representatives from member states and the E.U., chaired by the Commission. It will be responsible for coordinating the work of the national supervisory authorities and the Commission and providing advice and expertise to the Commission. In its advisory role, the EAIB will issue opinions, recommendations, and guidance on the AIA implementation and share best practices among member states.

The *governance system* of the AIA follows the ordoliberal principle of subsidiarity (Eucken, 1952/2004; Röpke, 1933/1965, 1942, 1944/1949, 1958/1961; Rüstow, 1955, 1957) and thus assigns essential roles to both member states and the Commission: The Commission oversees AIA monitoring and will establish a system for registering high-risk AI applications in a public E.U.-wide database. Member states, on the other hand, are responsible for establishing and designating national competent authorities, whose goal is to ensure the application and implementation of the AIA. Each state must also select one national supervisory authority amongst its national competent authorities (European Commission, 2021a). Those bodies act as a notifying and market surveillance authority, and they are tasked with assessing the operators' compliance with the AIA obligations and requirements for high-risk AI systems. Member states must also provide those authorities with adequate financial and human resources. The expertise must include an in-depth understanding of AI technologies, data and data computing, fundamental rights, health and safety risks, and knowledge of existing standards and legal requirements. The generated insights must be shared with the Commission and member states, and the national authorities can provide guidance and advice on the AIA implementation.

Besides the ex-ante conformity assessment, the AIA includes *post-market monitoring* and surveillance mechanisms. Providers, for example, are required to collect, document, and analyze data regarding the performance of high-risk AI systems and continuously evaluate the compliance of AI systems with the AIA. They must also investigate AI-related severe incidents and system malfunctioning and report these to the relevant authorities as a 'breach of obligations under Union law.' The national supervisory authorities are responsible for *market surveillance* and the control of AI systems. They are obliged to report to the Commission regularly on the outcomes of their activities and coordinate their efforts with other national authorities and the Union level. Noteworthy is that AI system providers are required to provide authorities with access to their data and documentation; under specific circumstances, authorities might even be granted access to the source code of AI systems. If corrective actions are insufficient to bring the system into AIA compliance, market surveillance authorities can require the provider or operator to withdraw the system from the market or recall it within a reasonable period. Organizations violating the AIA can be fined between EUR 10-30m or 2-6% of the global annual turnover (note that the Parliament wants to include fines of up to 7% [European Parliament, 2023c]).

At the time of writing, the revised AIA proposal has been approved by the Parliament (2023d), and negotiations with the Council are expected to start soon. Both institutions (and the Commission [Kazim et al., 2022]) must agree on a common text before the AIA can be adopted, which is expected to be in 2024.

## 4. Critical Ordoliberal Analysis

Given that we have discussed the *general* strengths[5] and weaknesses[6] of the AIA elsewhere (Wörsdörfer, 2023b), the following sections focus solely on the *ordoliberal* points of criticism.

---

[5] The *AIA's strengths* include its legally-binding, i.e., hard-law, character (Ebers, 2020, 2022), which marks a welcoming departure from existing soft-law AI ethics initiatives (Attard-Frost et al., 2022; EPRS, 2022a; Fjeld et al., 2020; Häußermann & Lütge, 2022; High-Level Expert Group on AI, 2019; Khanna, 2022; Leslie, 2019; Leslie et al., 2021; Mittelstadt, 2019; Rubenstein, 2021; White House, 2022), extra-territoriality and possible extension of the 'Brussels Effect' (Bradford, 2020; Floridi, 2021; Petit, 2020), ability to address data quality and discrimination risks (Hacker, 2021), and institutional innovations such as the EAIB and publicly accessible logs/database for AI systems (as an essential step in opening up black-box algorithms) (AlgorithmWatch, 2021). From a (revised) ordoliberal perspective, it is worth pointing out that the AIA attempts to ensure that AI technologies are 'ethically sound, legally acceptable, socially equitable, and environmentally sustainable, with a[n] [ordoliberal] vision of AI that seeks to support [i.e., serve] the economy, society, and the environment' (Floridi, 2021). Note that Röpke, Rüstow, and other ordoliberals also believed that the economy is embedded in a higher societal order – 'beyond supply and demand.' Its primary purpose is to serve the people and society, not vice versa. It is thus seen as a means to an end, not as an end in itself (the ordoliberal end in itself is the so-called 'vital situation' or 'private law society'). They also believed that the economy drains and erodes morality; 'moral reserves' thus need to be built outside the economy, i.e., in 'market-free' sectors and with the help of 'vital policy' (Böhm, 1966/1980; Röpke, 1942, 1950; 1950/1981; Rüstow, 1945/2001, 1960, 1961, 2001; Wörsdörfer, 2013).

[6] The *AIA's weaknesses* relate to its tendency to prioritize economic, business, and innovation over moral concerns (i.e., de-prioritization of human rights) (Almada & Petit, 2023; Castets-Renard & Besse, 2022; Floridi, 2021; Gstrein, 2022; Mazzini & Scalzo, 2022; Smuha et al., 2021; Wachter et al., 2021), the lack of a clear definition of AI systems (i.e., lack of scope) (AlgorithmWatch, 2021; EPRS, 2022a; Kazim et al., 2022; Mökander et al., 2022; Smuha et al., 2021), the flawed risk-based framework (i.e., incomplete list of prohibited AI systems and under-regulation of non-high-risk AI systems) (AlgorithmWatch, 2021; Biber, 2021; Ebers et al., 2021; EPRS, 2022a, 2022b; Gstrein, 2022; Mahler, 2022; Smuha et al., 2021; Stuurman & Lachaud, 2022), and the failure to adequately address the challenges posed by generative AI.

Ordoliberals are primarily concerned with the AIA's proposed governance structure. They specifically criticize its lack of …

- Effective enforcement,
- Oversight and control mechanisms,
- Procedural rights,
- Worker protection,
- Institutional clarity,
- Sufficient funding and staffing, and
- Consideration of sustainability issues.

1. *Lack of enforcement*: From a (revised) ordoliberal perspective, the fundamental problem with the AIA is the self-assessment and standardization process. If the draft proposal gets enacted as is, the governance process will rely significantly on self-monitoring by providers, especially in the form of the so-called 'ex-ante conformity assessment procedure.' The process is carried out by a provider, not an independent, i.e., external, third party (note that market surveillance by state authorities happens only ex-post, not ex-ante). As such, the proposed approach leaves the preliminary risk assessment, including identifying AI systems as high-risk to the provider. The latter thus possess a significant amount of discretionary leeway (Smuha et al., 2021), e.g., they can decide whether the used software is an AI system, whether the system may cause harm, and how to comply with the mandatory requirements of Title III, Chapter 2 of the AIA (Castets-Renard & Besse, 2022); furthermore, they can (theoretically) classify high-risk technologies as adhering to the rules using the self-assessment procedure (EPRS, 2022a). Ordoliberal critics thus fear that the regulations on prohibited and high-risk AI practices may prove ineffective given the absence of independent monitoring.

    Besides these extensive discretionary powers of AI providers, the AIA also grants sufficient leeway to standardization[7] and notified bodies[8]: The standardization process is one of the cornerstones of the AIA and is, therefore, essential for the conformity self-assessment

---

[7] *Standardization bodies* such as CEN and CENELEC are responsible for drafting harmonized voluntary technical standards, e.g., in the areas of electrical engineering.

[8] *Notified bodies* such as TÜV and other technical organizations are accredited by member states' notifying authorities and are responsible for assessing and verifying the conformity assessment procedure.

process. It requires providers to follow 'harmonized standards' developed by European standardization organizations such as CEN or CENELEC. Yet, delegating regulatory and standardization-setting powers to those bodies – i.e., transferring public rule-making competence to private associations – comes with various ethical issues, such as possible regulatory capture (Ebers, 2022). That is, intermediaries such as notified bodies might lack the required independence and impartiality from AI providers under assessment. The main reason is that they charge fees or receive a commission from AI providers and have thus an economic incentive to cater to their interests (Cefaliello & Kullmann, 2022). Besides these possible conflicts of interest, large AI providers, such as 'big tech,' possess significant market power, aggravating capture risks (Laux et al., 2023a). Overall, ordoliberals criticize that the current focus on self-assessment and corporate self-regulation is insufficient and might lead to ineffective or under-regulation. This risk of regulation circumvention is further exacerbated by the fact that the AIA, besides relying on self-assessments, also utilizes soft laws and codes of conduct for low-risk AI systems (Biber, 2021; Ebers et al., 2021; Stuurman & Lachaud, 2022).[9]

2. *Lack of democratic oversight and judicial control*: The delegation or outsourcing of rule-making power to standardization organizations (including notified bodies) might also lead to a lack of democratic oversight and the impossibility for relevant stakeholders, such as civil society organizations and consumer advocacy groups, to influence the drafting of standards as well as the judicial means to monitor and control them once they have been adopted (Castets-Renard & Besse, 2022; Ebers et al., 2021; Smuha et al., 2021).

    - *Lack of stakeholder consultation and participation*: Standardization processes and the related bodies often lack adequate involvement of relevant stakeholder groups, including consumer and civil society representatives (Castets-Renard & Besse, 2022). Especially consumer organizations often struggle to participate in such processes, e.g., due to a lack of financial resources, expertise, experience, or political representation (AlgorithmWatch,

---

[9] The AIA, for instance, encourages providers of *non-high-risk AI systems* to voluntarily apply the mandatory requirements for high-risk AI systems laid out in Title III. It also urges providers to *voluntarily* commit themselves – via codes of conduct – to environmental sustainability, accessibility for persons with disability, stakeholder participation, and team diversity.

2021; Gangadharan & Niklas, 2019; Micklitz & Gestel, 2013; Schepel, 2005; Veale & Zuiderveen Borgesius, 2021). Consequently, standardization processes tend to exclude certain stakeholder groups (this is especially the case for non-expert stakeholders and the public at large). As such, they often lack democratic input legitimacy.[10]

- *Power asymmetries*: Standards are usually developed by standard-setting organizations in which industry representatives exert significant influence, whereas the voices of civil society and consumer advocates are often overheard (note that, contrary to business organizations, civil society organizations often lack E.U.-level and agency representation). In the context of AI, there are also significant power imbalances between AI providers, on the one hand, and other relevant stakeholder groups and standardization and notified bodies, on the other (Castets-Renard & Besse, 2022).

- *Lack of transparency*: Little is known about the activities of standardization and notified bodies, mainly due to the frequent outsourcing of testing, inspection, and certification activities to third parties. Ordoliberal critics accuse those organizations of relying too much on corporate self-assessments instead of conducting independent and impartial monitoring (Veale & Zuiderveen Borgesius, 2021).

- *Consensus-finding problems*: AI regulation involves dealing with a variety of complex normative issues, such as defining or determining the acceptability and mitigation of pre-existing biases in training data (i.e., algorithmic fairness). Answering those and similar ethical questions means endorsing specific interpretations or theoretical approaches for normative concepts (e.g., equality, transparency, and dignity), or specifying acceptable or preferred trade-offs between competing interests (e.g., business vs. societal imperatives). It is questionable whether CEN or CENELEC will be able to resolve these politically sensitive issues, which the implementation of AI systems will inevitably raise. Therefore, researchers such as Laux et al. (2023b) predict that it will be hard to negotiate and find a consensus among different interest groups during the standardization process. They also point out that the above questions should be addressed by the Commission and Parliament[11] – i.e., by institutions that can be held democratically accountable and monitored by the judiciary – and not pushed down the line to standard-setting organizations, which are not democratically elected.

---

[10] Besides *input* legitimacy (i.e., a lack of stakeholder consultation and participation), those processes might also lack *throughput* (i.e., a lack of accountable and transparent processes) and *output* legitimacy (i.e., a lack of standard responsiveness, e.g., about the interests of affected stakeholder groups) (Laux et al., 2023b).

- *Techno-paternalism*: Critics also point out that the AIA equates trustworthiness with the acceptability of AI risks (hence distinguishing between acceptable/trustworthy and unacceptable/untrustworthy AI technologies). However, the acceptance and trustworthiness of technologies are primarily assessed through a *top-down* standardization process involving technology experts and conformity assessments carried out by AI providers (Laux et al., 2023a). From a (revised) ordoliberal point of view, such a risk regulation model should be avoided due to its paternalistic and technocratic nature and lack of bottom-up participation, i.e., its lack of public accountability, informed consent, and public communication (for an ordoliberal critique of techno-paternalism, see Klump & Wörsdörfer, 2015).

3. *Lack of procedural rights*: Individuals adversely impacted by AI systems have no right to complain to market surveillance authorities like they have the right to file a complaint and seek a judicial remedy against a supervisory authority under the GDPR. That is, the AIA creates no legal right to sue a provider, operator, or user for failures of AI systems (i.e., there is no way to contest and seek redress). Civil rights organizations, consumer advocacy groups, or other interested parties also lack rights similar to representative complaints possible under the existing data protection regime. Only those with obligations under the AIA can challenge regulators' decisions, not those whose fundamental (human) rights are negatively impacted. Given the lack of adequate complaint and judicial redress mechanisms, the AIA lacks a 'bottom-up force to hold (providers and) regulators accountable for weak AIA enforcement,' or to put it differently: the AIA lacks effective means of private enforcement (i.e., individual enforcement rights) to compensate for the ineffective public enforcement (Castets-Renard & Besse, 2022; Ebers et al., 2021; EPRS, 2022a; Smuha et al., 2021; Veale & Zuiderveen Borgesius, 2021).

---

[11] Note that the Parliament has no binding veto power over harmonized standards mandated by the Commission.

4. *Lack of worker protection*: AI systems are used for various purposes in human resource management, including but not limited to recruitment; performance management; task distribution, management, and evaluation; retention, rewards, and promotion; and disciplinary procedures. While 'AI as a (human resource) service' has clear benefits, such as enhancing operational efficiency and promoting occupational health and safety, it also raises ethical issues, such as erosion of privacy and data protection (i.e., AI bolsters the datafication of work and might lead to a so-called 'surveillance panopticon'), discrimination (i.e., biases and discriminatory effects of algorithmic management), and techno-paternalism and digital nudging (i.e., 'digital Taylorism') (Klump & Wörsdörfer, 2015). In addition, constant AI system monitoring at the workplace, e.g., via wearable technologies and other IoT devices, might lead to psychological harm and increased work stress and pressure (which, in turn, negatively impacts productivity). Most importantly, it potentially restricts workers' rights and negatively affects union activities (Cefaliello & Kullmann, 2022; EPRS, 2022b).

    According to ordoliberal critics, the AIA is less focused on protecting citizens' rights, including workers' rights, than the GDPR and other E.U. laws and regulations. It might thus contribute to a further undermining of employee rights. Especially problematic are AI-powered workplace monitoring and privacy intrusions, and that workers lack the right to object and file a complaint. Critics also point out that the list of prohibited AI practices at work is too narrow and that human oversight, assessment, and standardization processes do not (yet) require workers' – or social partners' – involvement. Overall, the AIA does not provide a sufficient counterbalance to the power of employers and AI providers, that is, adequate protection for workers exposed to AI systems and against the corresponding power imbalances (Cefaliello & Kullmann, 2022), as demanded by ordoliberalism.[12]

5. *Lack of institutional clarity*: The AIA implementation process involves seven(!) oversight institutions – i.e., market surveillance authorities, national supervisory authorities, notified authorities, notified bodies, conformity assessment bodies, post-market monitoring agencies, and the EAIB (Biber, 2021). From a (revised) ordoliberal perspective, having several institutions from various governance levels – e.g., national and E.U. – involved in the enforcement process is not necessarily problematic as it could be seen as a sign that the Commission follows an ordoliberal-inspired decentralized and horizontal approach – instead of a top-down one. However, having more than half a dozen institutions can hardly be justified, especially since their relationship remains unclear and vague. That is, the AIA lacks a proper coordination procedure, and the competencies of the respective agencies – and their relationship with each other – need to be clarified (EPRS, 2022a). Moreover, a lack of coordination (and streamlining) might also exacerbate ordoliberals' process policy concerns.[13] Lastly, it remains unclear how the supervision of the already existing governance regimes, such as the GDPR, DSA, and DMA, will work in practice with the newly established AI regulation, in general, and the EAIB, in particular (Gstrein, 2022).

---

[12] Note that Eucken and other ordoliberals saw unions as an essential counterweight to the power of employers (Eucken, 1952/2004, 1999).

As mentioned above, the EAIB – together with publicly accessible AI logs and database – has the potential to make a profound difference in terms of enforcement and monitoring. Yet, so far, no substantive powers have been conferred to the EAIB. It is also questionable whether it will be sufficiently independent of political influence – which is essential from a (revised) ordoliberal point of view – given that the AIA has allotted a predominant role to the Commission, which chairs the EAIB and has the right to veto the adoption of EAIB rules of procedure. Lastly, it remains to be seen what the cooperation mechanism between the EAIB, national authorities, and individuals affected by AI systems will look like (Ebers et al., 2021).

6. *Underfunding and understaffing*: Sufficient funding and staffing are essential for adequately enforcing the AIA, monitoring its compliance, and sanctioning non-compliance. The

---

[13] Process policy is rejected for several reasons: It is considered by Eucken and other ordoliberals as a form of 'privilege-granting policy.' It is mainly based on ad-hoc and case-by-case decisions and enables arbitrary and selective interventions in the economic 'game of catallaxy,' to use Hayek's (1973) term. It thus lacks two crucial features of an ordoliberal economic policy – predictability and long-term orientation. Most importantly, however, it opens the doors for special interest groups to exert influence on the legislative decision-making process: That is, process policy is more likely to be prone to the power of rent-seeking or lobbying groups – due to a more significant regulatory load and the existence of a higher discretionary leeway for decision-making. It thus goes hand in hand with a considerable lack of transparency as many debates and decisions take place behind closed doors – and a lack of accountability and democratic legitimacy – since interest groups represent only a fraction of society and are seldom directly and democratically elected (besides, process policy also tends to weaken or undermine constitutional checks and balances). In sum, this form of particularistic policy jeopardizes the nation's wealth – due to granting costly and exclusive privileges to special interest groups – and undermines personal freedom – due to the increased politico-economic powers of rent-seekers.

Commission estimates that the entire enforcement process of the AIA will only require between one and twenty-five extra full-time staff at the member state level (see AIA Impact Assessment, Annex 3 [European Commission, 2021c]). Veale and Zuiderveen Borgesius (2021) and others believe this to be insufficient for an effective AI governance regime.

A related problem is that standard-setting organizations and market surveillance authorities often lack the necessary technical expertise relative to AI providers. As such, many of those agencies often rely on market agents and outsource tasks and responsibilities to private actors – both of which exacerbate the already-discussed problems of lack of transparency, democratic accountability and legitimacy, and regulatory capture. Ordoliberal critics (Wörsdörfer, 2022b, 2023), therefore, demand that standardization bodies and government agencies must be better staffed and funded to avoid the above issues. This requires hiring additional staff – e.g., data analysts, AI specialists, and independent researchers – and (significantly) increasing the financial budgets of regulatory agencies (Almada & Petit, 2023).

7. *Lack of sustainability*: Eucken's Regulating Principles (1952/2004) demand the correction of negative external effects and the internalization of social costs, such as environmental pollution and other forms of ecological damage. Surprisingly, the AIA does not adequately address systemic sustainability risks created by AI systems, e.g., the significant energy (and water) consumption and corresponding greenhouse-gas emissions of data centers or the problem of electronic waste (EPRS, 2022a). There are also no direct references to climate change, ecological sustainability, environmental protection, and green or sustainable AI – except in the context of voluntary codes of conduct (Ebers et al., 2021). As Floridi (2021) and others have pointed out, the E.U. primarily follows an anthropocentric approach, prioritizing human needs and neglecting environmental concerns.

Figure 2 summarizes the AIA's institutional weaknesses – as seen from an ordoliberal perspective.

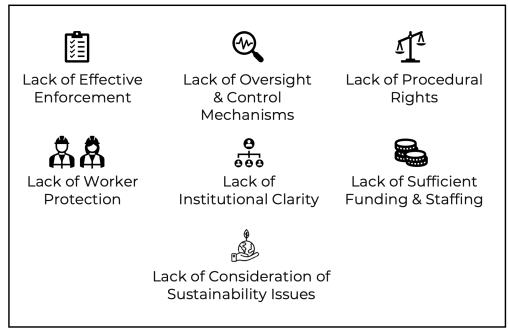

*Figure 2: Institutional Weaknesses of AIA*

## 5. Ordoliberal Reform Proposals

To address the previously described challenges and bring the AIA closer to an alignment with the ordoliberal ideals outlined in Section 2, several reform measures need to be taken. The following paragraphs focus, in particular, on introducing or strengthening (independent) conformity assessment procedures, democratic accountability and judicial oversight, redress and complaint mechanisms, worker protection, the AIA's governance structure, with a particular focus on the EAIB and AI database, funding and staffing, and sustainability requirements for AI providers. Ordoliberals believe that these proposals would help strengthen the AIA considerably.[14]

1. *Independent conformity assessment*: The current governance regime of the AIA rests on ex-ante product assessments and certifications (i.e., conformity assessment and CE marking) and ex-post market surveillance (i.e., post-market monitoring). While the latter is conducted by government agencies and, therefore, mandatory, the former is voluntary and self-regulatory in nature. From a (revised) ordoliberal point of view, this would need to change. I.e., there is a need for an independent conformity assessment carried out by an autonomous entity before market deployment (EPRS, 2022a). The goal must be to ensure that the risk assessment and certification process is more than a fig leaf: For this to work, the AIA needs to move beyond the currently flawed system of provider self-assessment and certification[15] towards mandatory third-party audits for all high-risk AI systems (Kop, 2021). The existing governance regime, which involves a significant degree of discretion for self-assessment and certification for AI providers and technical standardization bodies (Ebers et al., 2021; Gstrein, 2022), needs to be replaced with legally mandated external oversight by an independent regulatory agency with appropriate investigatory and enforcement powers (Cefaliello & Kullmann, 2022). Crucial in this regard is the safeguarding of the independence and impartiality of intermediaries. I.e., to avoid regulatory capture and conflicts of interest, the standardization and notified bodies need to be independent of the assessed AI provider.

---

[14] Note that Section 5 mirrors the previous section; i.e., the reform measures introduced in this section address the concerns raised in Section 4 in the exact same order (e.g., the paragraph on independent conformity assessment addresses the lack of enforcement concern in the previous section, and so on).

[15] According to AlgorithmWatch (2021), it is also hardly justifiable to leave the assessment of societal risks and impacts to corporate (i.e., for-profit) actors and their self-interests (i.e., profit and shareholder value maximization).

Having external oversight might thus also help in addressing some of the other ethical issues of the present standardization procedure, such as its opaqueness and lack of transparency, proneness to industry lobbying and corporate bias, lack of accessibility to and engagement with all relevant stakeholders, including civil society organizations, and technical standardization agencies' lack of ethical expertise, e.g., concerning fundamental rights. Ordoliberals recommend that the standardization procedure should ideally be conducted in a transparent, inclusive, and democratic manner (see below). This implies, among others, that the oversight mechanism for the conformity assessment process should include mandatory transparency requirements as well as auditing and third-party control mechanisms throughout the entire AI system lifecycle, which is especially important for self-learning systems (i.e., a conformity *re*-assessment is necessary if the system is subject to substantial modifications[16]) (AlgorithmWatch, 2021).

Laux et al. (2023b) go one step further and recommend that standard-setting organizations should develop AI standards that require 'ethical disclosure by default':

> "These standards will specify minimum technical testing, documentation, and public reporting requirements to shift ethical decision-making to local stakeholders and limit provider discretion in answering hard normative questions in the development of AI products and services [both of which are important from a revised ordoliberal perspective]. Rather than setting specific ethical requirements for trade-offs and thresholds, this approach would instead ensure all providers of AI systems meet a minimum harmonized standard for testing, reporting, and public participation. Ethical disclosure by default would exceed the reporting and participatory obligations currently proposed in the AIA draft regulation. Compliance would require AI providers to furnish relevant third parties with a standardized set of test results and documentation to enable local decision-makers to set normative requirements in a procedurally consistent way. Our proposed pathway is about putting the right information in the hands of the people with the legitimacy to make complex normative decisions at a local, context-sensitive level [following the ordoliberal principle of subsidiarity]" (Laux et al., 2023b, p. 7).

---

[16] Based on the notion of 'predetermined change,' anticipated AI system modifications currently do *not* trigger a new conformity assessment (Mazzini & Scalzo, 2022).

Those standards could utilize bias tests and de-biasing methods, fairness measures, open-source toolkits, transparency and explainability methods, model inspection methods, nutrition labels[17], impact assessments (e.g., privacy, algorithmic, and equality impact assessments[18]), and ethics review committees, to name a few.[19]

2. *Democratic accountability and judicial oversight*: From an ordoliberal standpoint, it is also crucial to strengthen democratic accountability and the judicial oversight mechanisms of the standardization procedure (EPRS, 2022a). What is needed is a meaningful engagement of all affected groups, including consumers and social partners (e.g., workers exposed to AI systems), and a public representation in the context of standardizing and certifying AI technologies. The overall (ordoliberal) goal is to ensure that those with less bargaining power are included and their voices are heard (EPRS, 2022b).

Currently, there is a considerable lack of democratic legitimacy (and accountability) due to the outsourcing of complex discussions around technical questions to technocratic experts and notified bodies. Those negotiations have far-reaching implications for human rights and other fundamental E.U. values and should thus not be left to non-democratically elected entities (Gstrein, 2022). Ebers et al. (2021), therefore, demand that policymakers take the necessary steps to improve the overall standardization process, including the structural and organizational framework of standardization organizations to facilitate an inclusive and democratic system[20] that provides for broad stakeholder engagement and dialogue and input on the development of technical standards.

---

[17] Stuurman and Lachaud (2022) argue for introducing mandatory *AI labeling schemes* for secure, responsible, and ethical AI systems. They claim that the current CE marking process (alone) is insufficient due to its lack of clarity, trust, monitoring, and transparency. Those labeling schemes could be similar to information or nutrition labels; that is, they would provide information about the goal of the AI system, the data collected and processed, and contact information to send queries and file complaints. The AI Ethics Label initiative suggests evaluating six dimensions of AI systems (transparency, accountability, privacy, justice/non-discrimination, reliability, and sustainability on a scale from A to G under a graphical design similar to the European energy efficiency label). The AI label would utilize ex-ante audits to validate the provided information, and such labels would need to be renewed regularly, given machine-learning progress (see for foundation, AI Transparency Institute, 2023; Thelisson et al., 2017).

[18] Some scholars request introducing mandatory algorithmic risk and human rights impact assessments for *any* AI-based application, not just for high-risk systems, as is currently planned. The risk levels could then be decided on a case-by-case basis, and systems either be banned or classified as high-risk. Such assessments would also promote the auditability and explainability of AI systems (AlgorithmWatch, 2021; EPRS, 2022b).

[19] A related concept – unlawfulness by default (Malgieri & Pasquale, 2022) – requires changing the burden of proof: i.e., the default is unlawfulness or unethicality, and AI providers have the burden to demonstrate that AI systems are not causing any harm, such as unfair/discriminatory decisions or inaccurate results, before they are marketed.

3. *Redress and complaint mechanisms*: Besides consultation and participation rights, ordoliberals also request the inclusion of explicit information rights and an effective complaint and redress mechanism (EPRS, 2022a): Currently, the AIA insufficiently considers the perspective of individuals and groups negatively affected by AI systems and possible remedies for such individuals and groups are not adequately addressed. AI subjects and other stakeholder groups, such as researchers, journalists, and representatives of civil society organizations, should, for instance, be able to access and retrieve relevant information, e.g., about the standardization and certification process as well as AI training data and data results. To safeguard this right to information, AlgorithmWatch (2021) demands the introduction of a legally binding data access framework. Furthermore, bearers of fundamental rights must have means to defend themselves if they feel they have been adversely impacted by AI systems or treated unlawfully. I.e., AI subjects must be able to legally challenge the outcomes of such systems. AlgorithmWatch (2021) thus demands easily accessible, affordable, and effective legal remedies and the introduction of individual and collective complaint and redress mechanisms (EPRS, 2022b; Smuha et al., 2021).

The ordoliberal principle of liability (Eucken, 1952/2004) also plays a role in the context of AI systems and the AIA: Individuals and groups negatively affected by AI technologies should be able to hold providers, developers, and users liable for any harm caused by their products. This, however, requires an accountability mechanism, which is currently lacking in the AIA (note that provisions on civil liability for damages caused by AI technologies are covered by the AI Liability Directive and the 'revised Product Liability Directive' but not by the AIA).

---

[20] Critics recommend making the standardization and risk-classifying process more transparent and inclusive, i.e., to have a better representation of stakeholder interests and counterbalance the adverse effects of private rulemaking (and the corresponding power imbalances between AI providers and other [civil society] stakeholders) (Ebers, 2022). It would require, among others, substantive information rights for affected individuals, adding public participation rights for citizens, e.g., regarding the decision to amend the list of high-risk systems, and ensuring that not only corporate and expert groups are involved in the standardization and risk-classifying process by actively involving organizations which represent public interests (Smuha et al., 2021).

4. *Worker protection*: AI systems tend to exacerbate the power imbalance between employers and workers. Ordoliberals thus demand better involvement and protection of workers, their representatives, and unions in using AI technologies at work (Cefaliello & Kullmann, 2022; EPRS, 2022b). This could be achieved in multiple ways: To begin with, more AI at-work systems could be classified as high-risk or prohibited (EPRS, 2022b). Workers should also be able to participate in management decisions regarding using AI tools in the workplace. Their voices and concerns should be heard, especially when technologies are introduced that might negatively impact their work experience (e.g., human oversight, assessment, and standardization processes should require workers' or social partners' involvement and representation). Workers should moreover have the right to object to the use of specific AI tools in the workplace and file complaints (Council of the European Union, 2022a, 2022b; European Parliament, 2023a, 2023b, 2023c; Smuha et al., 2021).[21] For instance, they could report issues with AI systems to an independent third-party agency or before the national supervisory authority. Unfortunately, the AIA does not include process requirements for participation and co-determination options for the use of AI systems (AlgorithmWatch, 2021); it thus fails to adequately protect workers exposed to AI systems and achieve a better power balance between employers and employees, as was envisioned by Eucken and other ordoliberals. Lastly, it might be worth exploring whether member states should be allowed to impose additional (i.e., better) employment-related requirements (EPRS, 2022b).

5. *Governance structure*: The ordoliberal principle of subsidiarity requires clarifying the relationship between member states and E.U. institutions and their respective roles and responsibilities. Ordoliberals tend to favor a decentralized instead of a centralized mode of regulation (Eucken, 1952/2004; Röpke, 1933/1965, 1942, 1944/1949, 1958/1961; Rüstow, 1955, 1957). Currently, however, the AIA mixes decentral(izing) elements with central(izing) ones. That is, on the one hand, it grants member states significant discretionary powers and influence, especially concerning post-market surveillance; on the other hand, however, E.U. institutions are responsible for the (self-regulatory) standardization and certification process, and the Commission is chairing the EAIB. According to ordoliberalism, what is needed is a clear(er) division of labor and clear-cut responsibilities (Gstrein, 2022). For instance, it is currently unclear whether – and if so, to what extent – member states can deviate from the AIA requirements, e.g., can they ban specific AI systems beyond the ones listed in the AIA and its annexes or can they introduce stricter or additional workplace-related requirements for such technologies in domestic law (AlgorithmWatch, 2021; Ebers et al., 2021)? In addition, the role of enforcement bodies – and their relationship with each other and E.U. institutions[22] – needs to be clarified (EPRS, 2022b).

---

[21] This is especially important in the context of (real-time) workplace monitoring, the potentially negative impacts of 'smart manufacturing' on labor markets (i.e., possible job losses), and other areas that might affect workers' rights (including freedom/autonomy rights).

Effective enforcement of the AIA also hinges on strong institutions and 'ordering powers' (Eucken, 1952/2004; Röpke, 1942, 1944/1949, 1950; Rüstow, 1955, 1957, 2001). The EAIB has the potential to be such a power and to strengthen AIA oversight and supervision. This, however, requires that it has the corresponding capacity, expertise (in both technology *and* fundamental rights), resources, and political independence. Ebers et al. (2021), however, question whether the EAIB possesses a sufficient degree of autonomy – given that the Commission chairs it; the researchers consequently demand more efforts to safeguard the board's independence (note that a similar proposition was recently made by the Council [2022a, 2022b] and Parliament [2023a]). It might also be worth exploring whether the EAIB's responsibilities could be expanded. Currently, the board serves in an advisory role, and experts claim that its tasks should be amended to include investigatory and regulatory powers; lastly, it could also be transformed into a stakeholder forum to overcome the previously mentioned issues of lack of consultation/ participation and stakeholder dialogue (Council of the European Union, 2022a, 2022b).

To ensure an adequate level of transparency – as demanded by ordoliberalism –, the E.U. database for AI systems should include not only high-risk systems but *all* forms of AI technologies. Moreover, it should list all AI systems used by private *and* public entities. The material provided to the public should include information regarding algorithmic risk and human rights impact assessment. This data should be available to researchers and those affected by AI systems in an easily understandable and accessible format (AlgorithmWatch, 2021; Smuha et al., 2021).

---

[22] E.g., it is crucial to provide some form of (minimum) harmonized implementation guidance to member states to prevent uneven or unreliable enforcement at the national level (Smuha et al., 2021).

6. *Funding and staffing of market surveillance authorities*: Besides the EAIB and AI database, national authorities must also be strengthened – both financially and expertise-wise. It is worth reiterating that the twenty-five full-time equivalent positions foreseen by the AIA for national supervisory authorities are insufficient and that additional financial and human resources need to be invested in the respective regulatory agencies to effectively implement the proposed AI regulation (AlgorithmWatch, 2021; EPRS, 2022b; Smuha et al., 2021).[23]

7. *Sustainability requirements*: Lastly, to better address the adverse external effects and environmental concerns of AI systems, ordoliberals also demand the inclusion of sustainability requirements for AI providers, e.g., obliging them to reduce the energy and water consumption of AI technologies and electronic waste (AlgorithmWatch, 2021; EPRS, 2022a), thereby moving towards green AI (Ebers et al., 2021). Ideally, those requirements should be mandatory and go beyond the existing codes of conduct.

A summary of ordoliberal-inspired reform proposals can be found in Figure 3.

---

[23] Moreover, E.U. policy and lawmakers should work towards harmonizing international AI standards and guidelines. Of particular importance in this regard is the transatlantic cooperation. Unified – and ideally global – standards for AI technologies would prevent regulatory gaps and 'forum shopping' (i.e., companies moving to countries with less regulatory burden and compliance costs) and would help in creating a level playing field with a minimum degree of legal certainty and planning security, as envisioned by ordoliberalism (Eucken, 1952/2004; Kop, 2021).

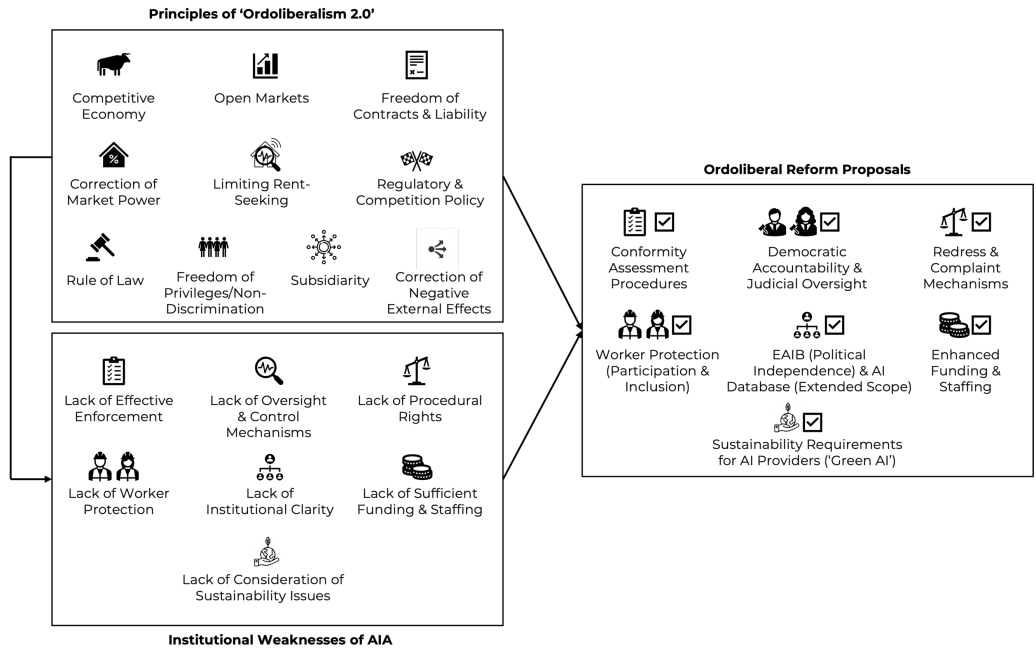

*Figure 3: Ordoliberal-Inspired Reform Proposals*

## 6. Concluding Remarks

The previous sections have analyzed the AIA from a distinct economic ethics perspective, i.e., 'ordoliberalism 2.0' (Wörsdörfer, 2020). Such a perspective has been absent in the academic literature so far. The article thus closes an important research gap and makes valuable contributions to the currently predominantly legal-political discourse on the AIA.

The article has, in particular, identified the AIA's institutional strengths and weaknesses, as seen from a (revised) ordoliberal point of view. The Act's main strengths include its mandatory and legally binding character, its extra-territoriality, and some of its institutional innovations, such as the EAIB (although its tasks and responsibilities need refinement and its political autonomy needs strengthening) and publicly available AI logs and databases. The most notable limitations of the AIA include its ineffective enforcement, the lack of proper oversight and control mechanisms, the missing procedural rights, the inadequate worker protection, the absence of sufficient funding and staffing, the lack of institutional clarity, and the insufficient consideration of AI sustainability. To address these challenges and align the AIA with 'ordoliberalism 2.0,' several ordoliberal-inspired reform measures have been identified, including but not limited to third-party conformity assessment, strengthened democratic accountability and judicial oversight, redress and complaint mechanisms, better worker protection, a reform of the AIA's governance structure (including the EAIB and the database for AI systems), enhanced funding and staffing, and the introduction of sustainability-related requirements for AI providers have been proposed. Those measures – in addition to others[24] – could help strengthen, i.e., harden, the AIA considerably.

As of the time of writing, the Parliament has approved the revised AIA proposal, and negotiations with the Council are expected to start soon. Both institutions (and the Commission) must agree on a common text before the AIA can get enacted. It remains to be seen how the final version will look and if it will incorporate at least some of the suggestions made in this article.

The challenges posed by generative AI and its underlying large-language models will likely necessitate additional AI regulation. The AIA thus needs to be revised and updated regularly. Future work in AI ethics and regulation needs to be vigilant of these developments and advise the Commission on how to amend the AIA and make it future-proof.

---

[24] AlgorithmWatch (2021) and others demand the explicit ban of all biometric mass surveillance technologies. According to the organization, the term 'real-time remote biometric identification systems' includes too many exceptions and ethical issues. For instance, it enables indiscriminate (i.e., arbitrarily or discriminatorily targeted) mass surveillance, which is incompatible with fundamental (human) rights and undermines key principles of rule-of-law societies. The organization also urges lawmakers to ensure that the ban applies to all public authorities, private actors that act on behalf of public authorities, and to post biometric identification systems – not only real-time. Lastly, the organization demands closing (some of) the AIA's loopholes, e.g., by removing the exceptions in Art. 5 (AlgorithmWatch, 2021). Other researchers recommend expanding the scope of the prohibition on social scoring to private actors, extending the ban on remote biometric identification systems in public spaces to non-law enforcement public actors, prohibiting the use of remote live biometric categorization systems in public places and the use of emotion recognition systems, and adding biometric categorization systems and emotion recognition systems to the list of high-risk systems (Smuha et al., 2021) (note that some of those concerns have been addressed by the Council and Parliament [see Section 3]). Most importantly, the Commission should be enabled to add AI technologies to the list of prohibited practices or high-risk systems. Here, it is crucial that the process of banning or adding high-risk categories is done in an inclusive, transparent, and democratic manner, that is, through robust consultation and stakeholder engagement, and that civil society representatives are heard. Also, all systems should be subject to prior independent conformity assessment control (Smuha et al., 2021).